\documentclass[12pt]{article}
\usepackage{amsmath,amssymb}
\newcommand  {\Ebar} {{\mbox{\rm$\mbox{I}\!\mbox{E}$}}}
\newcommand  {\Rbar} {{\mbox{\rm$\mbox{I}\!\mbox{R}$}}}

\newsavebox{\zzzbar}
\sbox{\zzzbar}
  {\setlength{\unitlength}{0.9em}
  \begin{picture}(0.6,0.7)
  \thinlines
  \put(0,0){\line(1,0){0.6}}
  \put(0,0.75){\line(1,0){0.575}}
  \multiput(0,0)(0.0125,0.025){30}{\rule{0.3pt}{0.3pt}}
  \multiput(0.2,0)(0.0125,0.025){30}{\rule{0.3pt}{0.3pt}}
  \put(0,0.75){\line(0,-1){0.15}}
  \put(0.015,0.75){\line(0,-1){0.1}}
  \put(0.03,0.75){\line(0,-1){0.075}}
  \put(0.045,0.75){\line(0,-1){0.05}}
  \put(0.05,0.75){\line(0,-1){0.025}}
  \put(0.6,0){\line(0,1){0.15}}
  \put(0.585,0){\line(0,1){0.1}}
  \put(0.57,0){\line(0,1){0.075}}
  \put(0.555,0){\line(0,1){0.05}}
  \put(0.55,0){\line(0,1){0.025}}
  \end{picture}}
\newcommand{\Zbar}{\mathord{\!{\usebox{\zzzbar}}}}

\newsavebox{\uuunit}
\sbox{\uuunit}
    {\setlength{\unitlength}{0.825em}
     \begin{picture}(0.6,0.7)
        \thinlines
        \put(0,0){\line(1,0){0.5}}
        \put(0.15,0){\line(0,1){0.7}}
        \put(0.35,0){\line(0,1){0.8}}
       \multiput(0.3,0.8)(-0.04,-0.02){12}{\rule{0.5pt}{0.5pt}}
     \end {picture}}

\newcommand{\QED}{{\hspace*{\fill}\rule{2mm}{2mm}\linebreak}}

\newcommand{\Z}{\Zbar}
\newcommand{\R}{\Rbar}

\newcommand{\E}{\Ebar}

\begin{document}

\title{{\bf  \Large FROM GLOBAL TO \\
LOCAL FLUCTUATION THEOREMS }} \vspace{10pt}
\author{Christian Maes\thanks{Instituut voor
    Theoretische Fysica, K.U. Leuven, Celestijnenlaan 200D, B-3001
    Leuven, Belgium - email:
    {\tt Christian.Maes@fys.kuleuven.ac.be } },
Frank Redig\thanks{T.U.Eindhoven. On leave from Instituut
    voor Theoretische Fysica, K.U. Leuven, Celestijnenlaan 200D, B-3001
    Leuven, Belgium - email: {\tt f.h.j.redig@tue.nl } } and
Michel Verschuere\thanks{ Instituut voor
    Theoretische Fysica, K.U. Leuven, Celestijnenlaan 200D, B-3001
    Leuven, Belgium - email:
    {\tt Michel.Verschuere@fys.kuleuven.ac.be } }
    }
\maketitle

\footnotesize
\begin{quote}
{\bf Abstract:} The Gallavotti-Cohen fluctuation theorem suggests a
general symmetry in the fluctuations of the entropy production, a basic
concept in the theory of irreversible processes, based on results in the
theory of strongly chaotic maps.
  We study this symmetry for some standard
models of nonequilibrium steady states.  We give a general strategy to
derive a {\it local} fluctuation theorem exploiting the Gibbsian features
of the stationary space-time distribution.  This is applied to spin flip
processes and to the asymmetric exclusion process.

\vspace{3pt}

\end{quote}
\normalsize

\footnotesize Dedicated in honor of Robert Minlos on the occasion of his
70th birthday. \normalsize

\vspace{12pt}

\section{Introduction}
A basic feature of equilibrium systems is that the restriction to a
subsystem is again in equilibrium and with respect to the same microscopic
interaction, for the same temperature, pressure and chemical potential.
We can imagine cutting out a much smaller but still macroscopic region
from our system and we will still find the same equilibrium state apart
from possible boundary effects.  Mathematically, this is expressed via the
DLR-equation stating that the local conditional probabilities of a Gibbs
measure coincide with the corresponding finite volume Gibbs measures.
This really amounts to the fact that, for Gibbs measures, the ratio of
probabilities for two different microscopic configurations that are
identical outside a finite volume, is given by the Boltzmann factor $\sim
\exp \beta \Delta H$ for relative Hamiltonian $\Delta H$ depending
continuously on the configuration far out.  In other words, relative
energies make sense and they can be written as a sum of sufficiently
local interaction potentials.  Important consequences are found in the
theory of equilibrium fluctuations and in the framework of the theory of
large deviations; Robert Minlos was among the very first to develop a
mathematical theory of this Gibbs formalism.  In particular in
\cite{min1}, the description of the thermodynamic limit of Gibbs measures
is given. We indeed usually have in mind here (very large)
spatially extended systems which are monitored locally.\\
  The idea that time does not enter this
equilibrium description is further stimulated from the fact that at least
in classical statistical mechanics, the momenta (entering only in the
kinetic energy part of the Hamiltonian) can be integrated out at once
from the partition function. Time enters already more explicitly in
equilibrium dissipative dynamics such as via Langevin equations (Markov
diffusion processes) or Glauber dynamics (for spin relaxation). Yet,
under the condition of detailed balance, the stationary dynamics is
microscopically reversible and the past cannot be distinguished from the
future.  The equilibrium steady state probability distribution on the
space-time histories still has a Gibbsian structure with as extra {\it
bonus} that the restriction to a spatial layer (at fixed time) is still
explicitly Gibbsian and in fact, the restriction of the dynamics to a
subregion still satisfies the detailed balance condition with respect to
it.  It was again Robert Minlos who was among the pioneers in the subject
of space-time Gibbs measures and who, with his great experience in cluster
expansion techniques and his love for field theory, saw the advantage of
the space-time approach in the construction of solutions to infinite
dimensional Markov diffusion processes, see e.g. \cite{min2}.\\  While
the condition of detailed balance reflects a symmetry (the space-time
distribution is time-reversal invariant), the fact that the space-time
distribution is Gibbsian (at least in some sense) does however not at all
depend on it. In other words, the fact itself that the spatio-temporal
probability distribution enjoys Gibbsianness is much more general and has
nothing to do with microscopic reversibility.  This can be checked
readily for probabilistic cellular automata, \cite{M1,M}, but, more
generally, it is the locality of the space-time interaction that does the
job. In the present paper, we will exploit this fact in going from a
global to a local fluctuation theorem for the entropy production in some
models of interacting particle systems.  At this moment, we need a second
introduction to write about the statistical mechanics of steady state
entropy production and how its fluctuations can give interesting
information about the response of the system to perturbations. We refer
to \cite{m0} for a recent impression.  As we will see and as introduced
in \cite{M},
 once the conceptual framework and the Gibbsian basis
of the fluctuation theorem is understood,  the transition from a global
to a local fluctuation theorem will be merely a technical matter.
Physically speaking however, it is much better for the obvious reason
that global fluctuations are far too improbable to be observed,
\cite{M,G3,GP}.

\subsection{Fluctuation theorem}
First observed in \cite{ecm} and later derived in \cite{gc1,gc2,Ru4}, the
Gallavotti-Cohen fluctuation theorem proves a symmetry in the fluctuations
in time of the phase space contraction rate for a class of dynamical
systems. The dynamics must obey certain conditions; it is a reversible
smooth dynamical system $\xi\mapsto \phi(\xi),\xi \in K$ on a phase space
$K$ that is in some sense bounded carrying only a finite number of
degrees of freedom (a compact and connected manifold). The transformation
$\phi$ is a diffeomorphism on $K$.  The resulting (discrete) time
evolution is obtained by iteration and the reversibility means that there
is a diffeomorphism $\theta$ on $K$ with $\theta^2=1$ and $\theta\circ
\phi \circ \theta = \phi^{-1}$.  It is assumed that the dynamical system
satisfies some technical (ergodic) condition: it is a transitive Anosov
system. This ensures that the system allows a Markov partition (and the
representation via some symbolic dynamics) and the existence of the SRB
measure $\rho$, an invariant measure with expectations
\begin{equation}\label{srb}
\rho(f) = \lim_N \frac 1{N} \sum_0^N f(\phi^n \xi)
\end{equation}
corresponding to time-averages for almost every randomly chosen initial
point $\xi\in K$ (i.e., for an absolutely continuous measure with respect
to the Riemann volume element $d\xi$ on $K$).  The change of variables
implied by the dynamics
  defines the Jacobian determinant $J$ and one writes
$\dot{S} \equiv - \ln J$. This is the phase space contraction rate which
Gallavotti-Cohen identify with the entropy production rate via the
following argument:
 Define the
(Shannon) entropy of a probability distribution $m(d\xi)=m(\xi)d\xi$ on
$K$ as
\begin{equation} S(m) = - \int d\xi
m(\xi) \ln m(\xi)
\end{equation}
With $m_n$ as density at time $n$, under the dynamics, the density at time
$n+1$ is
\begin{equation} m_{n+1}(v) =
\frac{m_n(\phi^{-1}v)}{J(\phi^{-1}v)}
\end{equation}
 and the change
in this entropy is therefore
\begin{equation}\label{enttime}
 S(m_N) - S(m_0) = \int d\xi m_0(\xi) \sum_0^{N-1} \ln J(\phi^n \xi)
\end{equation}
Dividing by $N$ and taking $N$ to infinity, the empirical probability
distribution approaches the SRB distribution $\rho$, as in (\ref{srb}).
Therefore, the time-averaged change in the entropy of the imagined
reservoir is $\rho(\dot{S})$, see also \cite{Ru1,Ru2,Ru4}. One further
assumes (and sometimes proves) dissipativity:
\begin{equation}\label{popo}
\rho(\dot{S}) > 0.
\end{equation}
 One is interested in the
fluctuations of
\begin{equation}\label{gccur}
\bar{s}_N(\xi) \equiv \frac 1 {\rho(\dot{S}) N} \sum_{-N/2}^{N/2}
\dot{S}(\phi^n(\xi)),
\end{equation}
in the state $\rho$. Informally, the fluctuation theorem then states that
$\bar{s}_N(\xi)$ has a distribution $\rho_N$ with respect to $\rho$ such
that
\begin{equation}\label{flucthm}
\lim_N \frac 1{N \rho(\dot{S})a}\ln \frac{\rho_N(a)}{\rho_N(-a)} = 1
\end{equation}
always.  In other words, the distribution of the time-averaged $\dot{S}$
over long time intervals satisfies some general symmetry property. A more
precise phrasing can be obtained via large deviation theory. For a
continuous time version (Anosov flows) we refer to \cite{Gen}.

The reason why we are interested in the fluctuation theorem is because
the established symmetry in fluctuations is very general and it may be
important for the construction of nonequilibrium statistical mechanics
beyond linear order perturbation theory. Now there are various proposals,
ideas and results but, at any rate,
 whatever the point of view, it is rather natural asking how to
establish a local version of the fluctuation theorem. The title of the
present paper refers to that problem with the understanding that local
refers to a space-time window within a much larger spatially and
temporally extended nonequilibrium system. This was already the subject
of \cite{M,G3}.  We want to understand how general the local fluctuation
theorem can be and what form it takes for some standard nonequilibrium
models.  It turns out that, as was explicitly pointed out in \cite{M}, it
is the Gibssian structure of the space-time distribution that allows a
local fluctuation theorem. This was already apparent from \cite{G3,M} but
here we add further systematization and explain and illustrate this local
version of the
fluctuation theorem.\\
One further question concerns the physical identification of the quantity
for which we are investigating the symmetry in the fluctuations. We will
call it entropy production.  This name already exists for a physical
quantity that appears in close to equilibrium thermodynamics, and indeed
we believe that our choice of words reflects a generalization. The basic
idea is that nonequilibrium steady states are not time-reversal invariant
and that the mean entropy production should give a measure of
discriminating between the original space-time distribution and its
time-reversal. That is the relative space-time entropy density. For the
variable entropy production, we must look up the source of the
time-reversal symmetry breaking in the space-time interaction. It turns
out that once it is recognized that the entropy production is the
antisymmetric part of the space-time interaction under time-reversal, the
symmetry in its local fluctuations (as expressed in the local fluctuation
theorem (LFT)) is
almost an immediate consequence of the Gibbsian structure.  This we will show.\\
Of course, the question remains how we wish to use the local fluctuation
theorem.  That is not the subject of the present paper but we refer to
\cite{G,G1,Ru4,M,m0,LeS,K} for some ideas.

\subsection{Example}
We sketch here the nature of a local versus global fluctuation theorem
via a simple model. We have in mind a $(1+1)$-dimensional Ising spin
system with formal Hamiltonian
\begin{equation}\label{hamt}
H(\sigma)=\sum_{x,t}\sigma_t(x)[\sigma_{t+1}(x)+b\sigma_{t+1}(x+1)]
\end{equation}
where we think $x\in\Z$ as the spatial coordinate and $t\in\Z$ as the
(discrete) time; $\sigma_t(x)=\pm 1,\ b\neq 0$.\\
Look at the function
\[
\bar{S}_{n,t}(\sigma)=b\sum^T_{t=-T}\sum^n_{x=-n}\sigma_t(x)[\sigma_{t-1}(x+1
)-\sigma_{t+1}(x+1)]\] of the spins in a space-time window parametrized
by $n,T>0$. We are interested in its fluctuations under the probability
laws
\begin{itemize}
\item $P_n$, the Gibbs measure on $\{-1,1\}^{\{-n,\ldots,n+1\}\times\Z}$ with
respect to the Hamiltonian
\[H_n(\sigma)=\sum_t\sum^n_{x=-n}\sigma_t(x)[\sigma_{t+1}(x)+b\sigma_{t+1}(x+1)
]\] and
\item $P$, any infinite volume Gibbs measure on $\{-1,1\}^{\Z^2}$ for the
Hamiltonian (\ref{hamt}). \end{itemize} In both cases we take the
counting measure as reference and set the inverse temperature $\beta=1$.
\\The difference is that $P_n$ is an Ising model on a one-dimensional strip
(finite spatial volume with infinite time-extension) and $P$ is the
corresponding model for infinite space-time volume.
\\We start with the statement of a global fluctuation theorem; that concerns
the law $P_n$. Consider the involution $\Theta_{n,T}$ by which all spins
inside the window
 $\Lambda_{n,T} =\{-n,\ldots,n+1\}\times \{-T-1,\ldots,T+1\}$ are reflected over the $t=0$ axis:
 $(\Theta_{n,T}\sigma)_t(x) = \sigma_{-t}(x)$ if $(x,t)\in \Lambda_{n,T}$ and remains unchanged otherwise.
Remark that $\bar{S}_{n,t}(\Theta_{n,T}\sigma)= -\bar{S}_{n,t}(\sigma)$
and upon writing $H_n(\Theta_{n,T}\sigma) - H_n(\sigma) =
\bar{S}_{n,t}(\sigma) - B_{n,T}(\sigma)$ we find, after a simple
calculation, that for every function $g$
\begin{equation}\label{localfl}
\int dP_n(\sigma) g(\bar{S}_{n,T}(\sigma))= \int
dP_n(\sigma)g(-\bar{S}_{n,T}(\sigma))e^{-\bar{S}_{n,T}(\sigma)+
B_{n,T}(\sigma)}
\end{equation}
with
$|B_{n,T}(\sigma)|\leq cn$.\\
As a result, for fixed $n$, for all functions $f$,
\begin{equation}\label{gloglo}
\lim_T\frac1{T} \ln|\frac{\int
dP_n(\sigma)f(\bar{S}_{n,T}(\sigma)/T)}{\int
dP_n(\sigma)f(-\bar{S}_{n,T}(\sigma)/T)e^{-\bar{S}_{n,T}(\sigma)}}|=0
\end{equation}
which implies the symmetry expressed in (\ref{flucthm}) with $N=T$. \\Now
to a local fluctuation theorem; that concerns the law $P$. A similar
calculation shows that $H(\Theta_{n,T}\sigma) - H(\sigma) =
\bar{S}_{n,t}(\sigma) - B_{n,T}(\sigma) - F_{n,T}(\sigma)$ with
\[
|F_{n,T}(\sigma)|\leq cT
\]
so that
 \[\int dP(\sigma)g(\bar{S}_{n,T}(\sigma))=\int
dP(\sigma)g(-\bar{S}_{n,T}(\sigma))e^{-\bar{S}_{n,T}(\sigma)+B_{n,T}(\sigma)
+ F_{n,T}(\sigma)}
\]
and we conclude that in both order of limits,
\begin{equation}\label{ditte}
\lim_{n,T}\frac1{nT}\ln|\frac{\int
dP(\sigma)f(\bar{S}_{n,T}(\sigma)/(nT))}{\int
dP(\sigma)f(-\bar{S}_{n,T}(\sigma)/(nT))e^{-\bar{S}_{n,T}(\sigma)}}|=0
\end{equation}
This is the same symmetry as in (\ref{flucthm}) but for the local
fluctuations in a spatially extended system. Of course, (\ref{ditte})
involves limits but the basic fact behind (\ref{ditte}) is that there is
a local function $R_{n,T} = H\circ\Theta_{n,T} - H $, antisymmetric under
the time-reversal $\Theta_{n,T}$ that preserves the {\it a priori}
reference measure,
 with
$|R_{n,T}(\sigma)-\bar{S}_{n,T}(\sigma)|\leq c_1n+c_2T$ for which
\[
\int dP(\sigma)g(R_{n,T})=\int dP(\sigma)g(-R_{n,T})e^{-R_{n,T}(\sigma)}
\]
which is an exact local fluctuation symmetry. Various things are lacking
from this example. Mathematically, things will be more complicated when
the $B_{n,T}$ or $F_{n,T}$ are not uniformly bounded or when time is not
discrete or when the space-time Hamiltonian (\ref{hamt}) is not local or
contains hard-core interactions. Physically, the example above carries no
interpretation of $\bar{S}_{n,T}$ as entropy production.
\subsection{Local fluctuation theorem}
The main theme of the present paper is
a general strategy to find a local fluctuation theorem for the entropy
production in a nonequilibrium steady state, in the context of stochastic
interacting particle systems. To get the idea we present
the result informally for a typical application. The details and
mathematically precise statements about this model are given in Section 4.
The model is a microscopic version of a reaction-diffusion system where
the reaction consists of the birth and death of particles on the sites of
a regular lattice and the diffusion part lets these particles hop to
nearest neighbor vacancies subject to an
external field.\\
Consider the square lattice $\Z^2$ to each site $i$ of which we assign a
variable $\eta(i) = 0,1$, meaning that site is empty or occupied by a
particle. The configuration $\eta$ can change in two ways: first, a
particle can be created or destroyed at lattice site $i$:
$\eta\rightarrow \eta^i$ where $\eta^i$ is identical to $\eta$ except
that the occupation at the site $i$ is flipped.  Secondly, a particle at
$i$ can hop to one of the four nearest neighbor sites $j$ under the
condition that $j$ is empty: $\eta\rightarrow \eta^{ij}$ where
$\eta^{ij}$ is the new configuration obtained by exchanging the
occupations at sites $i$ and $j$. We make a nonequilibrium dynamics by
adding an external field $E>0$ which introduces a bias for particle
hopping in a certain direction.
\\
In formula, first, a particle is destroyed or created at any given site
at fixed rates.  The transition from a configuration $\eta$ to the new
$\eta^i$ takes place at rate
\[
c(i,\eta) = \gamma_+ (1-\eta(i)) + \gamma_- \eta(i)
\]
where $\gamma_+$ is the rate for the transition $0\rightarrow 1$ and
$\gamma_-$ is the rate for $1\rightarrow 0$.
Secondly, the particles on the lattice undergo a diffusive motion. To be
specific, we choose a large square $V$ centered around the origin with
periodic boundary conditions and we first introduce hopping rates over a
nearest neighbor pair $\langle ij\rangle$ in the horizontal direction,
$i=(i_1,i_2), j=(i_1+1,i_2)$:
\[
c(i,j,\eta) = e^{E/2} \eta(i)(1-\eta(j)) + e^{-E/2} \eta(j)(1-\eta(i))
\]
The hopping rate in the vertical direction is constant (put $E=0$ in the
above if $j=(i_1,i_2\pm 1)$).  Taking $E$ large, we expect to see many
more jumps of particles to the right than to the left. In the absence of
reaction rates, that is for $\gamma_{\pm}=0$, we recover the so called
asymmetric exclusion process and particle number is strictly conserved.
More generally, the Master Equation is
\[
\frac{d\rho_t(\eta)}{dt} = \sum_i [\rho_t(\eta^i) c(i,\eta^i) -
\rho_t(\eta) c(i,\eta)] + \sum_{\langle ij\rangle} [\rho_t(\eta^{ij})
c(i,j,\eta^{ij}) - \rho_t(\eta)c(i,j,\eta)]
\]
For this model,
the stationary measure $\rho$
is the product
 measure
with uniform density equal to $\gamma_+ / (\gamma_- + \gamma_+)$
corresponding
to a chemical potential $\ln \gamma_+/\gamma_-$ of the particle reservoir.\\
For a fixed nearest neighbor pair $\langle ij\rangle$, with $j=
(i_1+1,i_2)$ to the right of $i$,
 the time-integrated microscopic
current over an interval $[-T,T]$ is
\[
J_T^{i} \equiv N_T^{i\rightarrow j} - N_T^{j\rightarrow i}
\]
with $N_T^{i\rightarrow j}$ the total number of particles that have passed
from site $i$ to site $j$.
We have the convention to take this current positive when the net number
of particles jumping to the right (i.e., in the direction of the external
field) is positive. Multiplying the sum of all the current contributions
in $V$ with the field $E$ we get
\[
W_{V,T}(\eta_s, s\in [-T,T]) \equiv E \sum_{i\in V} J_T^{i}(\eta_s, s\in
[-T,T])
\]
 which is a random variable representing the
work done on our system over the time-interval $[-T,T]$. Its  expectation
in the stationary state equals (up to a temperature factor) the
 expected
heat dissipated in the environment and is given by
\[
\langle W_{V,T}\rangle = 2T|V| \, E \sinh(E/2) \, \frac{\gamma_+
\gamma_-}{(\gamma_+ + \gamma_-)^2}
\]
If we now fix another square $\Lambda\subset V$ inside our large system,
then
\[
W_{\Lambda,T} \equiv E \sum_{i\in \Lambda: (i_1+1,i_2) \in \Lambda}
J_T^{i}
\]
is the random variable ``work done on the system in $\Lambda$ over the
time-interval $[-T,T]$". That constitutes the main contribution to the
local random variable ``entropy production in the space-time window
$\Lambda \times [-T,T]$".  Yet, this is only its bulk contribution.
 We have indeed only included in $W_{\Lambda,T}$ the
microscopic currents between the sites strictly inside $\Lambda$ while
particles will of course also hop in and out of $\Lambda$ via its
boundary. In other words, the region $V\setminus \Lambda$ acts as a
particle reservoir from which particles can enter or leave $\Lambda$.
That also contributes to the entropy production as, quite generally, the
change in entropy in the particle reservoir equals the number of particles
transferred to it, multiplied by its chemical potential.  Now usually,
this chemical potential is fixed and constant, i.e., not depending on
whatever happens in the system itself.  Here this is not the case.  It
suffices to imagine that almost all particles are in fact inside $\Lambda$
with therefore a low density of particles in $V\setminus\Lambda$.  As a
result, the {\it effective} chemical potential for creating or destroying
particles at the boundaries of $\Lambda$ will depend on time and on
whatever happened inside $\Lambda$ before that time.  Moreover this will
contribute to the nonequilibrium condition only for $E\neq 0$ because
only then will there be a different rate of leaving/entering $\Lambda$ at
the right versus the left vertical boundaries of $\Lambda$.  This is not
the case for the upper versus the lower boundaries but also there,
 even when there would not be a field strictly inside $\Lambda$, the
 dynamics
inside will be influenced by the field outside.  This is summarized in
the form of the second contribution to the time-integrated entropy
production and it is a boundary term:
\[
{\cal J}_{\partial \Lambda,T} \equiv R_\ell + R_r + R_u + R_d
\]
where the various terms correspond to the reactions taking place at the
left, right, upper and lower boundaries of the square $\Lambda$.  We will
not write all of them down explicitly but here is for example
\[
R_r \equiv \sum_{i\in \Lambda: j\in V\setminus \Lambda} \sum_{-T\leq t
\leq T} \eta_t(i)\ln \frac{\gamma_- + q_{\Lambda,t}(j,\eta,E)}{\gamma_- +
q_{\Lambda,t}(j,\eta,-E)} + (1-\eta_t(i))\ln \frac{\gamma_+ +
p_{\Lambda,t}(j,\eta,E)}{\gamma_+ + p_{\Lambda,t}(j,\eta,-E)}
\]
where $j=(i_1+1,i_2)$, the sum over times $t$ is over the times when a
particle is created or destroyed  at $i$, and the rates $p$ and $q$ are
given by
\[
q_{\Lambda,t}(j,\eta,E)\equiv
e^{E/2}\mbox{Prob}[\eta_t(j)=0|\eta_s(k),k\in \Lambda, s\in [-T,t]]
\]
and
\[
p_{\Lambda,t}(j,\eta,E)\equiv
e^{-E/2}\mbox{Prob}[\eta_t(j)=1|\eta_s(k),k\in \Lambda, s\in [-T,t]]
\]
where the probabilities refer to the steady state in $V$. In other words,
the external field does not only work on the particles in $\Lambda$ it
also creates a gradient in chemical potential (large at the left boundary
and smaller at the right) in $\Lambda$. The total random variable
``entropy production in $\Lambda$" now reads
\[
\bar{S}_{\Lambda,T} \equiv W_{\Lambda,T} + {\cal J}_{\partial \Lambda,T}
\]
 The result proved in Section 4 is the fluctuation theorem
symmetry for $\bar{S}_{\Lambda,T}$:
\begin{equation}\label{main}
\lim_{\Lambda,T}\lim_V \frac 1{|\Lambda|T} \ln \frac{\mbox{Prob}
[\frac{\bar{S}_{\Lambda,T}}{|\Lambda|T} = a]}
{\mbox{Prob}[\frac{\bar{S}_{\Lambda,T}}{|\Lambda|T} = -a]} = a
\end{equation}
uniformly in the $\gamma_{\pm}$.\\
One may wonder whether the work $W_{\Lambda,T}$
satisfies a similar fluctuation symmetry. That is (\ref{main}) with
$W_{\Lambda,T}$ replacing $\bar{S}_{\Lambda,T}$. It remains uncertain
however whether that is true uniformly in the values $\gamma_{\pm}\downarrow
0$ but, as we will show, it remains true whenever $\gamma_{\pm} \neq 0$.

The rest of our paper is organized as follows: in Section 2 we give a
general strategy to obtain LFT, which we apply in Section 3 for spinflip
processes and in Section 4 for the asymmetric exclusion process.

\section{Abstract setting}

We identify the essential mathematical structure, needed to pass from a
global to a local fluctuation theorem.  Our later specific illustrations
will then just be applications of the same theme.\\ We consider a
measurable space $(\Omega,{\cal F})$ on which two sequences of
probability measures $P_n$ and $P_n^r$.  Suppose that $\Theta_n$ is an
involution on $\Omega$ such that $P_n$ and $P_n\circ \Theta_n$ are
mutually absolutely continuous and the same for the pairs $P_n^r$ and
$P_n^r\circ \Theta_n$. We write
\[
R_n \equiv \ln \frac{dP_n}{dP_n \Theta_n}, F_n \equiv R_n + \ln
\frac{dP_n^r \Theta_n}{dP_n^r}
\]
 then, by definition, for all functions $f$,
\begin{equation}\label{core}
\int dP_n f(R_n) = \int dP_n e^{-R_n} f(-R_n)
\end{equation}
and
\begin{equation}\label{core1}
 \int dP_n^r f(R_n) =
\int dP_n^r e^{-R_n+F_n} f(-R_n)
\end{equation}
The identity (\ref{core}) expresses an exact symmetry in the fluctuations
of $R_n$ but should be compared with the global symmetry (\ref{localfl},
\ref{gloglo}). The next equality (\ref{core1}) is very similar but there
is the correction term $F_n$. To get rid of it (at least asymptotically
in $n$) we need extra assumptions. This will then yield the local
fluctuation theorem.  Before we give a general way of expressing these
assumptions, the reader may appreciate some more explication concerning
our choice of `global' versus 'local' as there is of course no natural
interpretation of this within the
proposed abstraction.\\
  As we will see in the next sections, we really
start from two measures $P$ and $P_n$ on $\Omega$ where $\Omega$ will be
the pathspace of an (infinite volume) interacting particle process on the
$d-$dimensional
 regular lattice $\Z^d$; $P$ will be an infinite volume
steady state measure (i.e., the path-space measure of a stationary
process over some time
 interval $[-T,T]$); $n$ will refer to a finite space-time volume
(corresponding to a sequence of
 cubes $\Lambda_n$ centered around the origin times the interval $[-T,T]$)
 and $\Theta_n$ will be time-reversal
on the space-time volume $\Lambda_n \times [-T,T]$. The process $P_n$
will be the path-space measure of the stationary interacting particle
process on this finite $\Lambda_n \times [-T,T]$.  $P_n^r$ is the
marginal distribution of the trajectories restricted to the space-time
window $\Lambda_n\times [-T,T]$ under $P$.
In the context of interacting particle systems, $P$ and $P_n$ will be
path-space measures of a Markovian process, whereas $P^r_n$ will be
non-Markovian. In the local fluctuation theorem it is attempted to
recover the global symmetry of $R_n$ under $P_n$ also in the restrictions
$P_n^r$ of $P$ to finite volumes $\Lambda_n$. Clearly then, what we need
is that the difference between $P_n$ and $P_n^r$ is a boundary term but
this is more or less implied by having
 our interacting particle systems enjoy Gibbsianness on
space-time.  Finally, the meaning of $R_n$ is that it gives, at least up
to space-time boundary terms,
 a statistical mechanical representation of the thermodynamic steady-state entropy
production.  We wish however to refer to \cite{M,M2,M3,M4,M5} for
explaining this.  Still, it should be kept in mind that the $B_n$
introduced in the following proposition will measure the difference
between the {\it true} entropy production (denoted there by
$\bar{S}_n$) and $R_n$.\\
There are in fact various strategies; we present two of them.

\noindent{\bf Proposition 2.1:} Let $B_n$ be a measurable function so
that $B_n\circ\Theta_n=-B_n$. Define $\bar{S}_n\equiv R_n+B_n$ and let
$(a_n)$ be a sequence of positive numbers tending to infinity with $n$.
Assume that $P_n$ and $P^r_n$ are mutually absolutely continuous and so
that
\begin{equation}\label{assmpt}
\lim_n\frac{1}{a_n}\ln\int
dP^r_n\left(\frac{dP_n}{dP^r_n}\right)^{\lambda_1} e^{\lambda_2 B_n}=0
\end{equation} for all $\lambda_1,\lambda_2\in\R$. Suppose that for all
$z\in \R$
\begin{equation} \label{helpme}
p(z) = \lim_n\frac1{a_n}\ln\int e^{-z\bar{S}_n}dP_n
\end{equation}
exist and is finite. Then, whenever
\begin{equation}
q(z) = \lim_n\frac1{a_n}\ln\int e^{-z\bar{S}_n}dP_n^r
\end{equation}
exists, then $p(z)=q(z)$ and $q(z) = q(1-z)$.


\noindent {\bf Remarks:}\\
1. The symmetry $q(z)=q(1-z)$ is dual to the symmetry as expressed in
(\ref{flucthm}).  Its Legendre transform $i(a)=\sup_z(-q(z)-za)$
satisfies $i(a)-i(-a)=-a$. If $\bar{S}_n$ satisfies a large deviation
principle under $P_n$, respectively  $P^r_n$, then $i(a)$ is the
corresponding rate function, and the symmetry $q(z)=q(1-z)$ is
equivalent with the large deviation symmetry $i(a)-i(-a)=-a$.\\
2. We will apply the strategy of Proposition 2.1 for obtaining a local
fluctuation theorem for spinflip processes in the next Section.

\noindent {\bf Proof of Proposition 2.1:} Since $B_n\circ\Theta_n=-B_n$,
in the same way as for (\ref{core}), we deduce that
\begin{equation}\label{leftright}
\int dP_n f(\bar{S}_n)= \int dP_n e^{-\bar{S}_n + B_n}f(-\bar{S}_n)
\end{equation}
Starting with the left hand side, for $f(s)=e^{-zs}$, by the H\"older
inequality, for $1/a+1/b = 1 = 1/v + 1/w$,
\begin{eqnarray}\label{erste}
&&\ln\int dP_n e^{-z\bar{S}_n} \leq \frac 1{a} \ln\int dP_n^r
(\frac{dP_n}{dP_n^r})^a + \frac 1{b}\ln\int dP_n^r e^{-bz\bar{S}_n}
\nonumber\\
&&\leq \frac 1{a} \ln\int dP_n^r (\frac{dP_n}{dP_n^r})^a +
\frac 1{bv}\ln\int dP_n e^{-bvz\bar{S}_n} + \nonumber \\
&&\frac 1{bw} \ln\int dP_n^r (\frac{dP_n^r}{dP_n})^{w-1}
\end{eqnarray}
Dividing this by $a_n$ and taking limits, we can use condition
(\ref{assmpt}) with $\lambda_2=0$ to get
\[
p(z)  \leq \frac{q(bz)}{b} \leq \frac{p(bvz)}{bv}
\]
Again by the H\"older inequality, both functions $p$ and $q$ are convex,
and hence continuous.  Therefore we can take the limit for $b,v\to1$ to
conclude that  $p(z)=q(z)$. The right hand side of (\ref{leftright}) can
be treated in the same way:
\begin{eqnarray}\label{right}
&&\ln\int dP_n e^{-(1-z)S_n + B_n} \leq \frac 1{a} \ln\int dP_n^r
(\frac{dP_n}{dP_n^r})^ae^{aB_n}+ \frac 1{b}\ln\int dP_n^r
e^{-b(1-z)\bar{S}_n}
\nonumber\\
&&\leq \frac 1{a} \ln\int dP_n^r (\frac{dP_n}{dP_n^r})^ae^{aB_n}+
\frac 1{bv}\ln\int dP_n  e^{-bv(1-z)\bar{S}_n +B_n}  + \nonumber\\
&&\frac 1{bw} \ln\int dP_n^r (\frac{dP_n^r}{dP_n})^{w-1} e^{-wB_n/v}
\end{eqnarray}
which, again after taking limits $n\uparrow +\infty$, and using
$B_n=\bar{S}_n - R_n$, gives
\[
p(z) =q(z) \leq \frac{q(b(1-z))}{b} \leq \frac{q(-bv(1-z)+1)}{bv}
\]
and we can take the limits $b,v\to1$ to get the desired $q(z)=q(1-z)$.
\QED \vspace{10pt}
\\{\bf Proposition 2.2:}
Let $B_n$ be a measurable function such that $B_n\circ\Theta_n=-B_n$ and
define $\bar{S}_n=R_n+B_n$. Let $(a_n)$ be a sequence of positive numbers
tending to infinity with $n$ so that for all $\lambda\in\R$
\begin{equation}\label{assmp}
\lim_n\frac1{a_n}\ln\int dP^r_ne^{\lambda(B_n+F_n)}=0
\end{equation}
Suppose that for all $z\in \R$
\begin{equation}
q(z) = \lim_n\frac1{a_n}\ln\int e^{-z\bar{S}_n}dP_n^r
\end{equation}
exists and is finite. Then, $q(z)=q(1-z)$.
\\{\bf Proof of Proposition 2.2:}
By definition of $F_n$, we have
\[\int dP^r_n f(\bar{S}_n)=\int dP^r_n e^{-\bar{S}_n+F_n+B_n}f(-\bar{S}_n)\]
We thus leave the left hand side and apply a similar chain of
inequalities to the right hand side as was used in the proof of
Proposition 2.1:
\begin{eqnarray}\label{rightr}
&&\ln\int dP_n^r e^{-(1-z)S_n + F_n + B_n} \leq \frac 1{a} \ln\int dP_n^r
e^{aF_n + aB_n}+ \frac 1{b}\ln\int dP_n^r e^{-b(1-z)\bar{S}_n}
\nonumber\\
&&\leq \frac 1{a} \ln\int dP_n^r e^{a(F_n + B_n)} + \frac 1{bv}\ln\int
dP_n^r e^{-bv(1-z)\bar{S}_n +F_n + B_n}
+\nonumber \\
&&\frac 1{bw} \ln\int dP_n^r e^{-w(F_n -B_n)/v}
\end{eqnarray}
We may thus again divide by $a_n$ and take limits first $n\uparrow
+\infty$ to reach
\[
q(z) \leq \frac{q(b(1-z))}{b} \leq \frac{q(-bv(1-z) +1)}{bv}
\]
By convexity we can take the limits $b,v\downarrow 1$ to obtain the
desired conclusion. \QED

\noindent {\bf Remarks:}\\
1. Of course, if it happens that $|F_n+B_n|/a_n\rightarrow 0$ uniformly,
then, for all positive functions $f$,
\[
\lim_n\frac1{a_n}\ln\frac{\int dP^r_n f(\bar{S}_n)}{\int
dP^r_ne^{-\bar{S}_n}f(-\bar{S}_n)}=0
\]
without further ado.\\
2. The difference between Proposition 2.1 and Proposition 2.2 is that in
the first we suppose that $P_n$ and $P_n^r$ are mutually absolutely
continuous while in the latter, we need that $P_n^r$ and $P_n^r\Theta_n$
are mutually absolutely continuous.  We will follow the second strategy
in Section 4 for the asymmetric exclusion
process.\\
3. The condition that the limits defining $p(z)$ and $q(z)$ exist is
natural in the context where we have a large deviation principle for
$\bar{S}_n$ under $P_n$ and $P^r_n$ resp. However if we define
$p^+,p^-,q^+,q^-$ by the corresponding limsup, resp.\ liminf, then we
still have convexity of $p^+,q^+$ (the limsups), but not necessarily of
$p^-,q^-$. We can still conclude however the equality $p^+ (z)= q^+ (z)$,
and $q^+(z)=q^+ (1-z)$.

\section{LFT for spinflip processes}
We start our study with the, for physical applications, less interesting
case of pure spinflip processes. For details on the construction of
spinflip
processes, we refer to \cite{LIGG}.\\
The configuration space is $K = \{+1,-1\}^{\Z^d}$ (spins on the
$d-$dimensional regular
 lattice) and the path space is
$\Omega= D(K,[-T,T])$ the set of right-continuous trajectories having left
limits, parametrized by time $t\in [-T,T], T>0$ and having values
$\omega_t\in K$. Our processes are specified in terms of spinflip rates
$c(x,\sigma), x\in \Z^d, \sigma\in K$ for which our first most important
assumption is that they are positive and bounded: there are constants
$b_1 > 0, b_2 < +\infty$ so that $b_1 < c(x,\sigma) < b_2$ for all
$x,\sigma$.
For convenience we
assume that $c(x,\sigma)$ only depends on the neighboring spins
$\sigma(y)$  with $|y-x|\leq 1$.
Thirdly, we assume the rates to be translation invariant: $c(x,\sigma)=
c(0,\tau_x \sigma )$.
Here and afterwards we  put $\Lambda_n= [-n,n]^d\cap\Z^d$
$\Theta_n$ denotes time-reversal on $\Lambda_n$ defined by $(\Theta_n
\omega)_t(x) \equiv \omega_{-t}(x)$ if $x\in\Lambda_n$, and $(\Theta_n
\omega)_t(x) \equiv\omega_{t}(x)$ if $x\notin\Lambda_n$. On the jump-times
we adapt $\Theta_n\omega$ so that it becomes right-continuous, and thus
obtain $\Theta_n$ as an involution on $\Omega $

We  define $\Lambda_n^\star \equiv \{x\in \Lambda_n, c(x,\sigma) =
c(x,\sigma') \mbox{ for all } \sigma,\sigma'\in K \mbox{ with }
\sigma(y)=\sigma'(y), y\in \Lambda_n\}$ for the subset of sites where the
spinflip rates do not depend on the configuration outside $\Lambda_n$.
\\
We first describe the sequence of processes $P_n$ corresponding to $P_n$
in the previous abstract setting.  For this we fix a boundary condition
$\eta\in K$ and we define spinflip rates
\begin{equation}
c_n(x,\sigma) \equiv I[x\in \Lambda_n]
c(x,\sigma_{\Lambda_n}\eta_{\Lambda_n^c}), x\in \Z^d,\sigma\in K
\end{equation}
where $I[\cdot]$ is the indicator function and
$\sigma_{\Lambda_n}\eta_{\Lambda_n^c} \in K$ coincides with $\sigma$ on
$\Lambda_n$ and equals $\eta$ on the complement $\Lambda_n^c\equiv \Z^d
\setminus \Lambda_n$. $P_n$ is the stationary process on $\Omega$  with
generator
\[
L_n f(\sigma) \equiv \sum_x c_n(x,\sigma) [f(\sigma^x) - f(\sigma)]
\]
corresponding to a spinflip process in $\Lambda_n$ with rates
$c(x,\sigma)$ and boundary condition $\eta$. We call the (unique)
stationary measure $\rho_n: \int d\rho_n L_nf =0$. We always assume that
for all $\sigma \in \{-1,+1\}^{\Lambda_n}, \rho_n(\sigma) \geq b_1
\exp[-b_2 |\Lambda_n|]$.
We can compute the density of $P_n$ with respect to $P_n\Theta_n$ via a
Girsanov formula for point processes, e.g. in \cite{B,LS}. For given
$\omega \in \Omega$ we let $N_s^x(\omega), s\in [-T,T], x\in \Z^d$ denote
the number of spinflips at $x$ up to time $s$; that is, $N_s^x(\omega)
\equiv |\{t\in [-T,s], \omega_{t^-}(x) = -\omega_t(x)\}|$; then,
\[
R_n \equiv \ln \frac{dP_n}{dP_n \Theta_n} = \sum_{x\in \Lambda_n}
\int_{-T}^T dN_s^x(\omega) \ln \frac{c_n(x,\omega_{s-})}{c_n(x,\omega_s)}
+ \ln \frac {\rho_n(\omega_{-T})}{\rho_n(\omega_T)}
\]
As a consequence, the distribution of $R_n$ as induced from $P_n$
satisfies
immediately the global fluctuation symmetry (\ref{core}).\\
We can also consider the observable
\[
\bar{S}_n \equiv  \sum_{x\in \Lambda_n^\star} \int_{-T}^T dN_s^x(\omega)
\ln \frac{c(x,\omega_{s-})}{c(x,\omega_s)}
\]
which is measurable inside $\Lambda_n$. This is not an arbitrary choice
but there is too little physics here to call $R_n$ or $\bar{S}_n$ the
entropy production; we will not elaborate on this. The difference $B_n
\equiv \bar{S}_n
- R_n$ is mainly a sum over $x\in \Lambda_n \setminus \Lambda_n^\star$.\\
The other process $P_n^r$ we need to look at is very similar but it is
 in general not Markovian. To define it, we take a stationary process $P$
on $\Omega$ and we take its restriction to $\Lambda_n$.  We write $\rho$
for the corresponding stationary measure and we let $\bar{\rho}_n$ be its
restriction to $\Lambda_n$. We always assume that for all $\sigma \in
\{-1,+1\}^{\Lambda_n}, \bar{\rho}_n(\sigma) \geq a_1 \exp[-a_2
|\Lambda_n|]$. Being more explicit, we let $P$ be an infinite volume
stationary process with formal generator
\[
Lf(\sigma) \equiv \sum_x c(x,\sigma) [f(\sigma^x) - f(\sigma)]
\]
and put $P^r_n$ the unique path-space measure such that
\begin{enumerate}
\item The distribution of $\{ \omega_t (x):x\in\Lambda_n, t\in [-T,T]\}$
under $P^r_n$ and $P$ coincide.
\item Under $P^r_n$, $\omega_t (x) = \eta (x)$ for all $x\not\in\Lambda_n$,
$t\geq 0$.
\end{enumerate}
\noindent {\bf Theorem 3.1} [LFT for spinflip processes] For all $z\in
\R$,
\[
\lim_{n,T} \frac 1{n^dT} \ln \frac{\int dP e^{-z\bar{S}_n}}{  \int dP
e^{-(1-z)\bar{S}_n}} = 0
\]
\noindent {\bf Proof of Theorem 3.1:}\\
Even though $P_n^r$ is not Markovian (in general), it remains a jump
process and the jump-intensities can be computed from the original
spinflip rates. In order to have a Gibbsian structure these intensities
must be the same in the bulk of $\Lambda_n$ as they were for the infinite
volume process $P$. As the rates are local, the process $P_n^r$
restricted to $\Lambda_n$ indeed has the same intensities as the process
$P_n$ except at the sites of the boundary $\Lambda_n \setminus
\Lambda_n^\star$. This is a consequence of the following generally stated

\noindent {\bf Lemma 3.2:}\\
Suppose $N_t$ is a point process with intensity $c_s$, i.e.,
$M_t=N_t-\int^t_0c_s\ ds$ is a martingale for the filtration
$\mathcal{F}_t$. Suppose that $\mathcal{F}'_t\subset\mathcal{F}_t$ is a
subfiltration of $\mathcal{F}_t$, and define
\begin{equation}
N'_t=\Ebar[N_t|\mathcal{F}'_t]
\end{equation}
Then $N'_t$ is a point process with intensity
\begin{equation}
c'_s=\Ebar[c_s|\mathcal{F}'_s]
\end{equation}
{\bf Proof of Lemma 3.2:}\\
It is easy to see that $M'_t=N'_t-\Ebar(\int^t_0
c_sds|\mathcal{F}'_t)=\Ebar[M_t|\mathcal{F}'_t]$ is a $\mathcal{F}'_t$
martingale.
Hence, it suffices to show that
\begin{equation}
B_t=\Ebar[\int^t_0c_sds|\mathcal{F}'_t]-\int^t_0\Ebar[c_s|\mathcal{F}'_s]ds
\end{equation}
is a $\mathcal{F}'_t$-martingale. This is a consequence of the following
equalities:
\begin{eqnarray}
&&\Ebar[B_t|\mathcal{F}'_s]\nonumber\\
&&=B_s+\Ebar\left[\int^t_s(c_r-\Ebar[c_r|\mathcal{F}'_r])dr|\mathcal{F}'_s\right]\nonumber\\
&&=B_s+\Ebar[\int^t_s
c_rdr|\mathcal{F}'_s]-\Ebar\left[\int^t_s\Ebar[c_r|\mathcal{F}'_r]|\mathcal{F}'_s\right]\nonumber\\
&&=B_s
\end{eqnarray}
\QED Therefore, the rates of the restricted process on $\Lambda_n$ are
given by
\[\tilde{c}_s(x,\omega)=\Ebar[c(x,\sigma_s)|\sigma_\tau(y)=\omega_\tau(y),-T\leq
\tau\leq s,y\in\Lambda_n]\]
where the expectation is with respect to $P$.\\
Or, for all $x\in\Lambda_n$, $N^x_t (\omega)-\int_0^t
\tilde{c}_s(x,\omega)$ is a martingale under $P^r_n$. As a consequence,
$\tilde{c}_s(x,\omega)=c(x,\sigma)$ when
$x\in\Lambda^\star_n$ and $\omega_s=\sigma$. \\
Just as for the pair $P_n, P_n\Theta_n $, the absolutely continuity of
$P_n^r\theta_n$ with respect to $P_n^r$ and vice versa is guaranteed by
the positivity of the spinflip rates inside $\Lambda_n$. We are thus
ready to apply Proposition 2.1.
We must first verify the corresponding assumption (\ref{assmpt}). We find
\begin{equation}
B_n=\bar{S}_n-R_n=-\sum_{x\in\Lambda_n\setminus\Lambda^\star_n}\int
dN^y_s(\omega) \ln \frac{c_n(x,\omega_{s^-})}{c_n(x,\omega_s)}+
\ln\frac{\rho_n(\omega_T)}{\rho_n(\omega_{-T})}\end{equation} and
\begin{equation}
\ln\frac{dP^r_n}{dP_n}=
\ln\frac{\bar{\rho}(\omega_{-T})}{\rho_n(\omega_{-T})}+
\sum_{x\in\Lambda_n\setminus\Lambda^\star_n} \int
dN^x_s(\omega)\ln\frac{\tilde{c}_s(x,\omega)}{c_n(x,\omega_s)}-
\int^T_{-T}ds[\tilde{c}_s(x,\omega)-c_n(x,\omega_s)]
\end{equation}
Clearly, both $|B_n|$ and $|\ln dP^r_n/dP_n|$
are bounded by $c_1
N([-T,T],\Lambda_n\setminus\Lambda^\star_n)+c_2|\Lambda_n| + c_3T
|\Lambda_n\setminus\Lambda^\star_n|$
 for
some constants $c_1,c_2,c_3<\infty$,
 where $N([-T,T],\Lambda_n\setminus\Lambda^\star_n )$ is the
number of spinflips that have occurred in the space-time window
$[-T,T]\times(\Lambda_n\setminus\Lambda^\star_n)$. It remains thus to
show for all $\lambda$
\begin{eqnarray*}
&&\lim_{n\uparrow\infty}\frac{1}{n^d}\ln\int dP^r_n e^{\lambda
N([-T,T],\Lambda_n\setminus\Lambda^*_n)}\\
&&=\lim_{n\uparrow\infty}\frac{1}{n^d}\ln\int dP^0 \frac{dP^r_n}{dP^0_n}
e^{\lambda N([-T,T],\Lambda_n\setminus\Lambda^*_n)} =0 \end{eqnarray*}
where we inserted the reference process $P^0$ (and its restriction
$P^0_n$ to $\Lambda_n$) corresponding to the product process of
independent spinflips (rate 1).  In particular, \begin{equation*}
\lim_{n\uparrow\infty}\frac{1}{n^d}\ln\int dP^0 e^{\lambda
N([-T,T],\Lambda_n\setminus\Lambda^*_n)}=
\lim_{n\uparrow\infty}\frac{|\Lambda_n\setminus\Lambda^*_n|}{n^d}2T(e^{\lambda}
-1)=0 \end{equation*} and we can apply the same argument as in
Proposition 2.1. Finally, the condition (\ref{helpme}) of Proposition 2.1
is
a consequence of the large deviation results of \cite{daipra}.
\QED

\section{LFT for the asymmetric exclusion process}
The configuration space is now $K=\{0,1\}^{\Z^2}$ (occupation variables
 on the 2-dimensional regular lattice)
and the pathspace $\Omega=D(K,[-T,T])$ is essentially unchanged from that
in the previous section. For $\eta\in K$, $\eta(x)=1,0$ indicates the
presence, respectively absence of a particle at the site $x\in \Z^d$.
This hopping dynamics will be modeled by an asymmetric exclusion process.
This is a bulk driven diffusive lattice gas.  The hopping rates for
vertical ($v$) and horizontal bonds ($h$)
 depend
on the direction in the following way:
\begin{eqnarray*}\label{vh}
c^v(x,\eta) \equiv \frac 1{2} [\eta(x)(1-\eta(x+e_2))+\eta(x+e_2)(1-\eta(x))]\\
c^{h,E}(x,\eta) \equiv \frac 1{2} [e^{E/2}\eta(x)(1-\eta(x+e_1))+e^{-E/2}
\eta(x+e_1)(1-\eta(x))]
\end{eqnarray*}
where $e_1, e_2$ are the unit vectors in the positive horizontal and
vertical
direction.\\
In addition, for the moment, we allow for the possibility of particle
creation and destruction. We put the birth/death rate $c(x,\eta) \equiv
\epsilon$ independent of the
configuration $\eta$ and the site $x$.\\
The formal Markov generator $L$ to the infinite volume process is then
found as the sum
\[
Lf(\eta) \equiv \epsilon \sum_x [f(\eta^x) - f(\eta)] + \sum_{\langle
xy\rangle} c(x,y,\eta)[f(\eta^{xy}) - f(\eta)]
\]
where $\eta^x$ is the new configuration after changing the occupation at
$x$, $\eta^{xy}$ is the new configuration after exchanging the
occupations at $x$ and $y$ and $c(x,y,\eta)$ is given by (\ref{vh}) for
nearest neighbors $x,y = x \pm e_i$. We can allow more general
reaction-diffusion processes (e.g. with extra interaction, speed change,
etc.) but we will stick here to this choice. What is simpler here is that
the Bernoulli measure $\rho$ with density 1/2 is a non-reversible
stationary measure.  The corresponding pathspace measure over the
time-interval $[-T,T]$ is $P=P^E$ and we put $P^E_n$ the process
restricted to the finite square $\Lambda_n$
 This $P^E_n$ will now play the role of $P^r_n$ of Section 2.
 For
a given trajectory $\omega\in\Omega$ we let $N^{xy}_s(\omega),
s\in[-T,T]$ be the number of hopping times where the occupation at the
nearest neighbor
sites $\langle xy \rangle$ was exchanged.
 Since this model has a clear physical interpretation
we can
define the variable entropy production in $\Lambda_n$.

The first contribution comes from the work done by the external field
\begin{equation}\label{618}
\bar{W}_n\equiv E\sum_{\langle xy\rangle,y=x+e_1 \in
\Lambda_n}\int^T_{-T}dN^{xy}_s(\omega)
[\omega_{s^-}(x)(1-\omega_{s^-}(y))-\omega_{s^-}(y)(1-\omega_{s^-}(x))]
\end{equation}
This is of the form field ($E$) times current.
There is a second contribution from differences in the reaction rates at
the boundary: particles enter or leave at different rates at the various
boundaries; this contribution is present even in the case where no
external field is applied inside $\Lambda_n$:
\begin{eqnarray}\label{verg2}
{\cal J}_n(\omega) &\equiv & \sum_{x\in \partial \Lambda}\sum_{y=x\pm
e_1\in \Lambda_n^c} \int_{-T}^T dN_s^x(\omega)
[\omega_{s^-}(x)\ln\frac{2\epsilon + \lambda_y^E(\omega,s)e^{E_j/2}}
{2\epsilon + \lambda_y^{-E}(\omega,s)e^{-E_j/2}}\nonumber\\ &&+
(1-\omega_{s^-}(x))\ln\frac{2\epsilon + \kappa^E_{y}(\omega,s)e^{-E_j/2}}
{2\epsilon + \kappa^{-E}_{y}(\omega,s)e^{E_j/2}}] \nonumber\\
&&+ \sum_{x\in \partial \Lambda}\sum_{y=x\pm e_2\in \Lambda_n^c}
\int_{-T}^T dN_s^x(\omega) [\omega_{s^-}(x)\ln\frac{2\epsilon +
\lambda_y^E(\omega,s)} {2\epsilon + \lambda_y^{-E}(\omega,s)}\nonumber\\
&&+ (1-\omega_{s^-}(x))\ln\frac{2\epsilon + \kappa^E_{y}(\omega,s)}
{2\epsilon + \kappa^{-E}_{y}(\omega,s)}]
\end{eqnarray}
where the second sum is over all (external) neighbors $y$ of $x$ and
$\partial \Lambda_n$ is the interior boundary of $\Lambda_n$. Here, the
additional rates are
\[
\lambda^E_{y}(\eta,t) \equiv \E^E[1-\eta_t(y)|\eta_s(z), z\in \Lambda_n,
s\in [-T,t]]
\]
and
\[
\kappa^E_{y}(\eta,t) \equiv \E^E[\eta_t(y)|\eta_s(z), z\in \Lambda_n,
s\in [-T,t]]
\]
for $E_j=\pm E$ if $y=x\pm e_1, E_j=0$ if $y=x\pm e_2$ and the
expectations are in the process $P=P^E$. The variable entropy production
is put
\[
{\bar S}_n = {\bar W}_n + {\cal J}_n
\]
The symmetry in the fluctuations of $\bar{S}_n$ is given by

\noindent {\bf Theorem 4.1} [LFT for the aymmetric exclusion process] For
all $\epsilon$ (including $\epsilon=0$), for all $z\in R$,
\[
\lim_{n,T} \frac 1{n^2T} \ln \frac{\int dP e^{-z \bar{S}_n}}{  \int dP
e^{-(1-z)\bar{S}_n}} = 0
\]
\noindent {\bf Proof of Theorem 4.1} We start by noting that for the
time-reversal $\Theta_n$,
\[
P^E_n \Theta_n = P^{-E}_n
\]
Obviously then, for a function $f$ measurable in $\Lambda_n \times
[-T,T]$,
\[
\int dP f(\omega) = \int dP^E_n f(\omega) = \int dP^{E}_n
\frac{dP_n^{-E}}{dP_n^{E}} f(\Theta_n \omega)
\]
and we must investigate express the density $dP_n^{E}/dP_n^{-E}$ via a
Girsanov formula.  This is the strategy of Proposition 2.2. The Girsanov
formula gives
\begin{equation}\label{lang}
\ln \frac{dP_n^{E}}{dP_n^{-E}} = \bar{S}_n + F_n
\end{equation}
with the following correction term:
\begin{eqnarray}\label{f0}
&& F_n(\omega) \equiv \sinh(E/2) \sum_{x,y=x + e_1\in
\Lambda_n}\int_{-T}^T ds
 [\omega_s(y)(1-\omega_s(x)) -
\omega_s(x)(1-\omega_s(y))] \nonumber\\ &&+ \sum_{x\in \partial
 \Lambda} \sum_{y=x\pm e_1,x\pm e_2\in \Lambda_n^c} \int_{-T}^T ds
 [\omega_s(x) [\lambda_y^E(\omega,s)) -
 \lambda_y^{-E}(\omega,s)]\nonumber\\
 &&+ (1-\omega_s(x))[\kappa_y^E(\omega,s)) - \kappa_y^{-E}(\omega,s)]
\end{eqnarray}
Now, $|F_n| \leq c |\partial \Lambda| T$ because the (first) bulk term in
(\ref{f0}) telescopes to a boundary term.  We can thus apply Remark 1
after Proposition 2.2 to finish the proof.\QED

Next, we investigate whether the variable work $W_n$ of (\ref{618})
itself satisfies the same local symmetry as the entropy production.

\noindent {\bf Theorem 4.2} [LFT for the work done] For all $\epsilon>0$,
for all $z\in \R$,
\[
\lim_{n,T} \frac 1{n^2T} \ln \frac{\int dP e^{-z\bar{W}_n}}{  \int dP
e^{-(1-z)\bar{W}_n}} = 0
\]
\noindent {\bf Proof of Theorem 4.2} Clearly, since $\epsilon > 0$,
$|{\cal J}_n| \leq c N([-T,T],\Lambda_n\setminus\Lambda^*_n)$, that is
bounded, up to a constant, by the number of flips in the trajectory on
sites $x\in \Lambda_n\setminus\Lambda^*_n$, for times $t\in[-T,T]$.  We
can therefore verify condition (\ref{assmp}) in the same way as we did
for Theorem 3.1.  Finally, the large deviation results of \cite{daipra}
remain valid for $\epsilon>0$, so that we can finish the proof along the
lines of Proposition 2.2.\QED

\section{Remarks}

\begin{enumerate}
\item
It is clear from the preceding analysis that the reasons for having a
global or local fluctuation theorem do not in any way depend on the
$\Theta_n$ being time-reversal.  Thus, the same results will be
reproduced in exactly the same form for any other involution.  Of course,
the symmetry breaking part in the pathspace action functional will be the
variable for which the fluctuation symmetry holds (replacing entropy
production corresponding to time-reversal symmetry breaking).  As an
example, if for a spinflip process, the rates are not even under a global
spinflip (by the presence of a bias or magnetic field), then a local
fluctuation theorem will be established  for the variable magnetization.
Furthermore, we may consider the composition of two or more involutions
--- in this way, we could e.g. obtain a local fluctuation theorem for the
odd part (under spinflip) of the variable entropy production.  Finally,
we can even go beyond  the case of involutions and consider instead the
generators of the symmetry group for the unperturbed dynamics.  In this
case, the precise form of the fluctuation symmetry is not preserved but
its modification presents no real problem.
\item
We restricted our discussion to interacting particle systems where the
evolution is Markovian. Within the Gibbsian space-time picture, this
means that the interaction is
``nearest neighbor" in the time direction (the jump
intensity at time $t$ depends only on the configuration at time $t^-$).
However, this restriction is not at all necessary. If the jump
intensities are local in space and bounded from above and from below,
then we can still apply the Girsanov formula for point processes to
obtain the local fluctuation theorem from the global fluctuation theorem.
\end{enumerate}

\end{document}